\RequirePackage{fix-cm}
\documentclass[twocolumn,epjc3]{svjour3}  
\smartqed  
\RequirePackage{graphicx}
\RequirePackage{amssymb}
\RequirePackage{lipsum}
\RequirePackage{amsmath}
\RequirePackage{graphicx}
\RequirePackage{subcaption}
\RequirePackage{hyphenat}
\RequirePackage{microtype}
\RequirePackage{mathtools}
\usepackage[utf8]{inputenc}
\usepackage[T1]{fontenc}
\RequirePackage{xcolor}
\RequirePackage{arydshln} 
\RequirePackage{lineno}
\RequirePackage{cite}
\RequirePackage{slashed}
\RequirePackage{comment}
\RequirePackage{orcidlink}

\journalname{Eur. Phys. J. C}
\begin{document}

\title{Comprehensive Constraints on ALP Couplings from future $e^+e^-$ Colliders, Muon $g-2$, Thermal Dark Matter and Higgs Measurements}


\author{Pramod Sharma\thanksref{e1,addr1,addr2}
        \and
        Soham Singh\thanksref{e2,addr2,addr3}
        \and Mukesh Kumar\thanksref{e3,addr2}\,\orcidlink{0000-0003-3681-1588}
        \and Ashok Goyal\thanksref{e4,addr2,addr4}
}

\thankstext{e1}{e-mail: pramodsharma.iiser@gmail.com} 
\thankstext{e2}{e-mail: sohamsingh931@gmail.com} 
\thankstext{e3}{e-mail: mukesh.kumar@cern.ch} 
\thankstext{e4}{e-mail: agoyal45@yahoo.com}



\institute{Indian Institute of Science Education and Research, Knowledge City, Sector 81, S. A. S. Nagar, Manauli PO 140306, Punjab, India. \label{addr1}
           \and
           School of Physics and Institute for Collider Particle Physics, University of the Witwatersrand, Johannesburg, Wits 2050, South Africa. \label{addr2}
           \and
           Department of Physics \& Astronomy, National Institute of Technology, Rourkela 769 008, India.\label{addr3}
           \and
           Department of Physics \& Astrophysics, University of Delhi, Delhi 110 007, India.\label{addr4}
}

\date{Received: date / Accepted: date}

\maketitle

\begin{abstract}
In this article, we present projected 95\% C.L. limits on Axion-Like Particle (ALP) couplings from ALP production at a future $e^+e^-$ collider operating at $\sqrt{s} = 250~\text{GeV}$ with integrated luminosity $L = 0.5~\text{ab}^{-1}$. We constrain the effective couplings $g_{\gamma\gamma}$, $g_{Z\gamma}$, $g_{ZZ}$, and $g_{WW}$ over the ALP mass range $20~\text{GeV} \leq m_a \leq 100~\text{GeV}$, finding projected bounds at the level of $\mathcal{O}(10^{-1})~\text{TeV}^{-1}$ for $g_{\gamma\gamma}/f_a$. Given that the latest muon anomalous magnetic moment measurement ($\Delta a_\mu$) shows no statistically significant deviation from the Standard Model prediction, we reinterpret the ALP contributions to $\Delta a_\mu$ as a stringent consistency requirement. We then derive the corresponding allowed regions for $g_{\gamma\gamma}$ and the ALP--muon coupling $C_{\mu\mu}$, and apply them to a fermionic dark matter scenario in which the relic density depends on both the dark matter mass $m_\chi$ and $m_a$. The same parameter space is further constrained by Higgs signal strength measurements through $h \to \gamma\gamma$ and $h \to Z\gamma$. A comparative analysis with existing experimental and theoretical bounds highlights the complementarity of $\Delta a_\mu$, dark matter, and Higgs observables in restricting ALP couplings, demonstrating that even in the absence of a $\Delta a_\mu$ anomaly, these constraints provide essential guidance for viable ALP parameter space.
\end{abstract}

\section{Introduction}
\label{intro}

Axion-like particles (ALPs) have emerged as a compelling framework for addressing multiple unresolved questions in particle physics. In addition to their origin as pseudo-Nambu-Goldstone bosons of broken global symmetries~\cite{Georgi:1986df,Chikashige:1980qk}, they offer phenomenological richness across various energy scales. For instance, ALPs have been invoked to help resolve the long-standing discrepancy in the muon's anomalous magnetic moment $\Delta a_\mu$~\cite{Marciano:2016yhf}, and when coupled to fermionic dark matter (DM), can enable thermal freeze-out consistent with the observed relic density~\cite{Izaguirre:2016dfi}. Furthermore, loop-induced ALP couplings to electroweak Standard Model (SM) gauge bosons can modify Higgs-boson ($h$) decay rates, potentially accounting for mild excesses observed in $h \to \gamma\gamma$ and $h \to Z\gamma$ channels at the Large Hadron Collider (LHC)~\cite{ATLAS:2020qcv,CMS:2022ahq,ATLAS:2023yqk}.

In this article, we intend to present and compare constraints on ALP couplings to electroweak gauge boson pairs ($V = W^\pm, Z, \gamma$) in the mass range $20~\mathrm{GeV} \le m_a \leq 100~\mathrm{GeV}$. Our primary focus is on results from a dedicated analysis at future $e^+e^-$ colliders, which we place in context alongside constraints derived from the $\Delta a_\mu$, dark matter (DM) relic density,\footnote{In Ref.~\cite{Holst:2021lzm}, the authors explore a new parameter space within a gauged $L_\mu - L_\tau$ extension of the SM that accommodates both the $\Delta a_\mu$ and a viable MeV-scale dark matter candidate.} and recent Higgs signal strength measurements. Finally, we compare our results with existing bounds from previous studies, including recent projections from $e^-p$ collider scenarios~\cite{Bauer:2017ris,Brivio:2017ije,Mosala:2023sse}, thereby highlighting the complementarity and discovery potential of the $e^+e^-$ collider program in probing ALP scenarios.

After electroweak symmetry breaking, the ALP ($a$) interacts with a fermionic DM candidate $\chi$ via a renormalizable coupling $C_{a\chi}$, while its interactions with the SM charged leptons $\ell$ and electroweak gauge bosons are described by dimension-5 operators in the effective Lagrangian~\cite{Georgi:1986df,Bauer:2017ris,Brivio:2017ije,Bauer:2018uxu}:
\begin{equation}
\begin{split}
\mathcal{L}_{\rm eff}^{\rm int} \supset & -i \frac{C_{a\chi}}{2} \partial_\mu a \bar{\chi} \gamma^\mu \gamma^5 \chi - \sum_{\ell=e,\mu,\tau} i \frac{C_{\ell\ell}}{2} \frac{\partial_{\mu} a}{f_a} \bar{\ell} \gamma^\mu \gamma^5 \ell \\
& + e^2 \frac{a}{f_a} g_{\gamma\gamma} F_{\mu\nu} \tilde{F}^{\mu\nu}  + \frac{2e^2}{c_w s_w} \frac{a}{f_a} g_{Z\gamma} F_{\mu\nu} \tilde{Z}^{\mu\nu} \\
& + \frac{e^2}{c_w^2 s_w^2} \frac{a}{f_a} g_{ZZ} Z_{\mu\nu} \tilde{Z}^{\mu\nu} + \frac{e^2}{s_w^2} \frac{a}{f_a} g_{WW} W_{\mu\nu} \tilde{W}^{\mu\nu},
\end{split}
\end{equation}
where $c_w$ and $s_w$ are the cosine and the sine of the Weinberg mixing angle $\theta_W$, respectively. For all studies in this work, the scale parameter is fixed at $f_a = 1$~TeV. To perform our analysis, we implemented this Lagrangian in \texttt{FeynRules}~\cite{Alloul:2013bka}, built the required model files, and used them in our simulations.

\begin{figure*}[t]
\centering
\subfloat[]{\includegraphics[width=.50\textwidth]{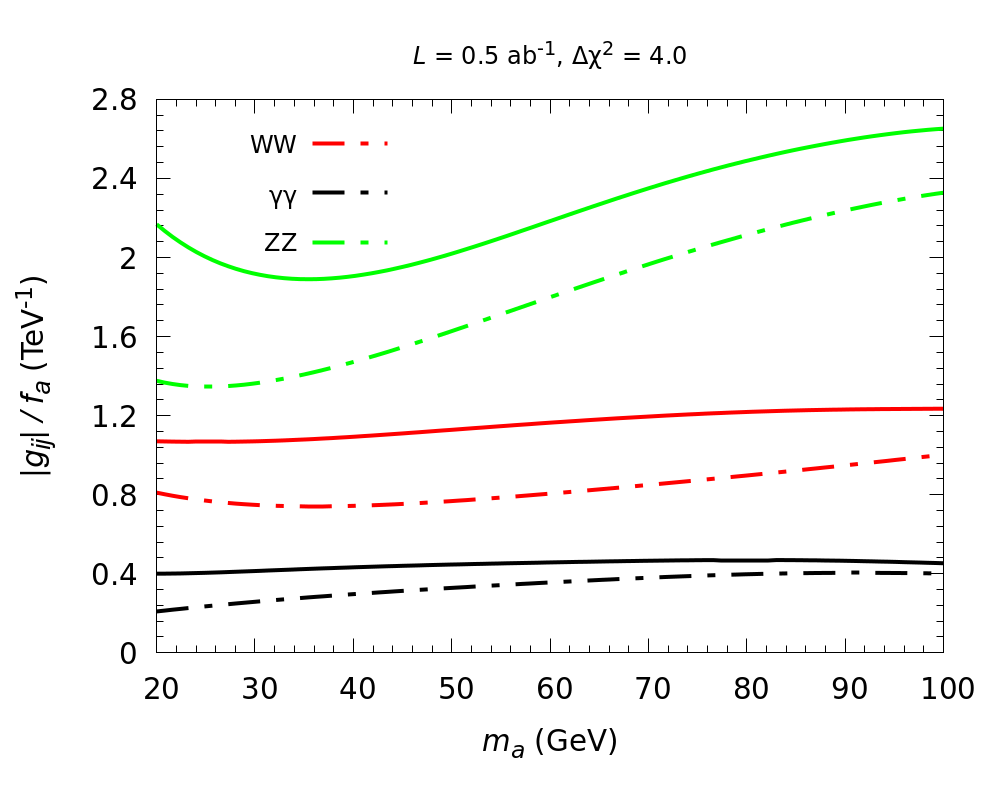} 
\label{fig:one_param}}
\subfloat[]{\includegraphics[width=.50\linewidth]{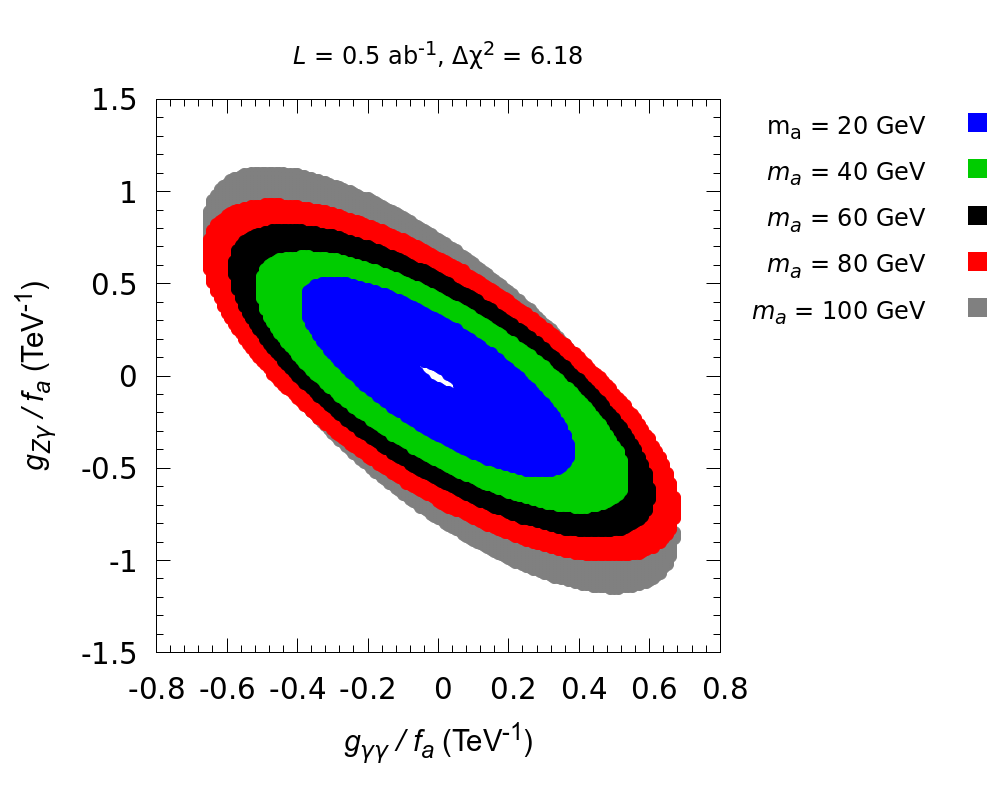}
\label{fig:two_param}}
\caption{The 95\% C.L. exclusion contours for $t$-channel ALP production at an $e^+e^-$ collider with an integrated luminosity of $L = 0.5$~ab$^{-1}$, obtained from a $\chi^2$ analysis, are shown: (a) in the $g_{ii}/f_a$–$m_a$ plane using one-bin (solid lines) and multi-bin (dashed lines) approaches, and (b) in the $g_{Z\gamma}$–$g_{\gamma\gamma}$ plane for a two-parameter multi-bin analysis, assuming a fixed coupling of $g_{ZZ} = 0.1$. In panel (b), the colored regions correspond to the allowed parameter space, while the regions outside are excluded at 95\% C.L.}
\label{fig:gij}
\end{figure*}

\begin{figure}[t]
\centering
\includegraphics[width=.48\textwidth]{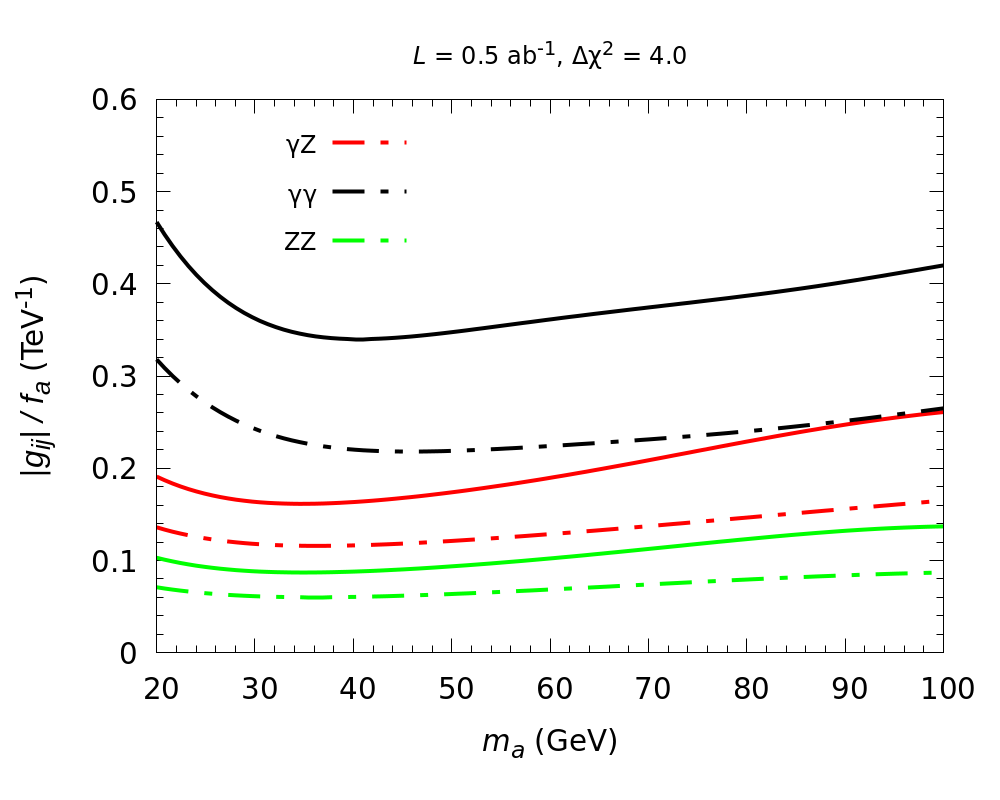}
    \caption{The 95\% C.L. exclusion contours for Higgsstrahlung ALP production at an $e^+e^-$ collider with an integrated luminosity of $L = 0.5$~ab$^{-1}$, obtained from a $\chi^2$ analysis. Results are shown in the $g_{ii}/f_a$–$m_a$ plane using one-bin (solid lines) and multi-bin (dashed lines) approaches.}
    \label{fig:chi_sChan}
\end{figure}

\section{Analysis \& Results}
\label{analysis}
In this section, we present our analysis to derive constraints on the couplings $g_{ij}$ ($i,j = W^\pm, Z, \gamma$) as a function of $m_a$ from the observables at the $e^+e^-$ collider, including $\Delta a_\mu$, relic density, and the Higgs signal strengths reported in~\cite{ATLAS:2020qcv,CMS:2022ahq,ATLAS:2023yqk}.

\subsection{\texorpdfstring{$e^+e^-$}-collider:}
\label{const-ee}
To estimate constraints on the couplings $g_{ij}$ at a future $e^+e^-$ collider, we consider ALP production through both $t$-channel (fusion) and $s$-channel (Higgsstrahlung) processes.

The $t$-channel ALP production follows via charged-current (\texttt{CC}) ($WW$-fusion) and neutral-current (\texttt{NC}) processes ($\gamma\gamma$, $ZZ$, or $Z\gamma$-fusion), in association with scattered $\nu_e \bar{\nu_e}$ and $e^+e^-$, respectively. In particular, we focus on the decay $a \to \gamma\gamma$, with the branching ratio (Br) taken as a function of $m_a$. In this case, individual couplings are probed by setting $g_{ii}=1$ while keeping all other couplings at zero. However, for $g_{Z\gamma}$, we set $g_{WW} = 0$, while keeping $g_{\gamma\gamma} \neq 0$ and $g_{ZZ} \neq 0$, so that interference between the $\gamma\gamma$, $ZZ$, and $Z\gamma$-fusion channels is taken into account. Note that $\text{Br}(a \to \gamma\gamma)$ is the dominant decay mode for ALP masses below $m_a < 100$~GeV~\cite{Mosala:2023sse}. The \texttt{CC} and \texttt{NC} signal and corresponding background events, with final states $\nu_e \tilde{\nu_e} \gamma\gamma$ and $e^+e^- \gamma\gamma$, respectively, with center-of-mass energy $\sqrt{s} = 250$~GeV, are generated using the Monte Carlo event generator package \texttt{MadGraph5}~\cite{Alwall:2011uj}. Further showering, fragmentation, and hadronization are performed using \texttt{Pythia8}~\cite{Sjostrand:2007gs}, while detector-level simulation is carried out with \texttt{Delphes}~\cite{deFavereau:2013fsa}, based on the proposed CEPC detector design~\cite{Chen:2017yel,CEPCStudyGroup:2023quu}. The factorization and normalization scales are set to dynamic scale. For this study, $e^\pm$ polarization is taken to be $\pm 80$\%. 

\begin{figure*}[t]
    \centering
    \subfloat[\label{fig1a}]{\includegraphics[width=0.25\linewidth]{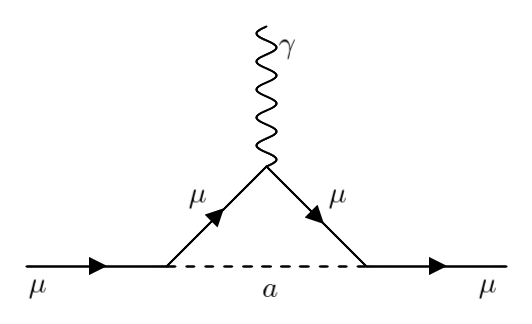}}
    \subfloat[\label{fig1b}]{\includegraphics[width=0.25\linewidth]{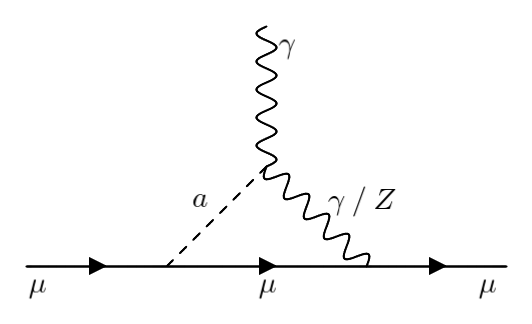}}
    \subfloat[\label{fig1c}]{\includegraphics[width=0.25\linewidth]{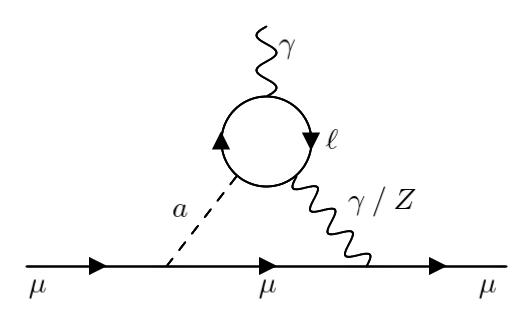}}
    \subfloat[\label{fig1d}]{\includegraphics[width=0.25\linewidth]{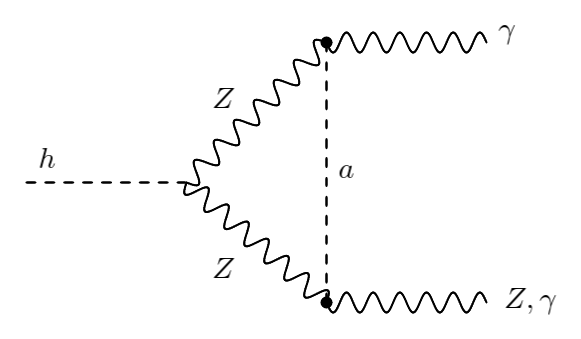}}
    \caption{Leading (a), (b) one- and (c) two-loop Feynman diagrams contributing to the muon anomalous magnetic moment, involving an axion-like particle (ALP) and its interactions with SM leptons and neutral electroweak gauge bosons. (d) Feynman diagrams involving ALP couplings $g_{Z \gamma}$ and $g_{ZZ}$ contributing to the loop-induced decays $h \rightarrow Z \gamma$ and $h \rightarrow \gamma \gamma$.}
    \label{fig:mug}
\end{figure*}
\begin{figure*}[t]
\centering
    \subfloat[\label{fig2a}]{\includegraphics[width=0.5\linewidth]{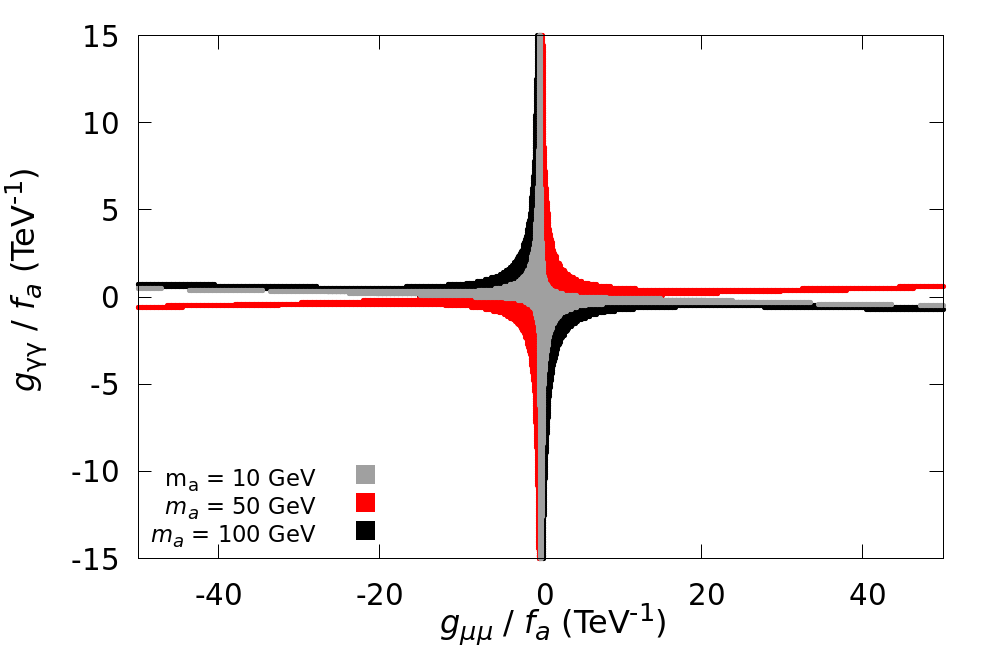}}
    \subfloat[\label{dis-yy}]
{\includegraphics[width=.5\linewidth]{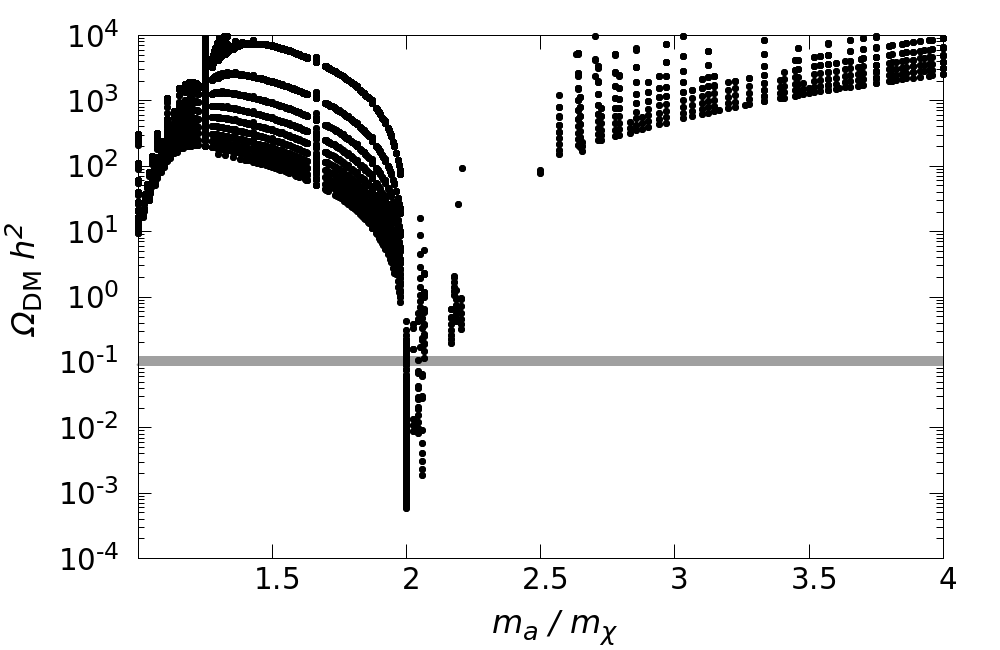}}
\caption{The $1\sigma$ contours consistent with the $\Delta a_\mu$ anomaly. (a) Allowed regions in the $(g_{\gamma\gamma}/f_a,\, g_{\mu\mu}/f_a)$ plane for $m_a = 10$, 50, and 100~GeV. The cutoff scale is set to 1~TeV. (b) Relic density as a function of $m_a/m_\chi$ (scatter plot) for couplings that resolve the $\Delta a_\mu$ anomaly. The observed relic density ($2\sigma$ band) is indicated by the horizontal shaded region, intersecting the curve at two viable values of $m_a$.}
\label{fig:g-2}
\end{figure*}

At the generation level, nominal selection criteria are imposed, with transverse momentum and rapidity cuts of $p_T^{e^\pm, \gamma} > 10$~GeV and $|\eta^{e^\pm, \gamma}| < 2.5$, respectively. To enhance signal sensitivity against dominant backgrounds, additional analysis-level selections are employed. The di-photon invariant mass is constrained within $|m_{\gamma\gamma} - m_a| < 7.5$~GeV. For all channels, the leading photon must satisfy $p_T^{\gamma} > 20$~GeV for $30~\mathrm{GeV} < m_a < 50~\mathrm{GeV}$, and $p_T^{\gamma} > 30$~GeV for $50~\mathrm{GeV} < m_a \leq 100~\mathrm{GeV}$. For the $\gamma\gamma$ and $Z\gamma$ fusion channels, forward-backward rapidity cuts on the scattered $e^\pm$ are applied: $-2.5 < \eta^{e^-} < 0$ and $0 < \eta^{e^+} < 2.5$, while for the $ZZ$ fusion channel, the selection $-1 < \eta^{e^-} < 2.5$ and $-2.5 < \eta^{e^+} < 1$ is imposed. In the $WW$ fusion channel, a central rapidity requirement of $-1 < \eta^{\gamma} < 1$ is used, along with a missing transverse energy threshold of $E_T^{\rm miss} > 20$~GeV.

Further, to constrain the ALP-gauge couplings $g_{ij}$, we perform a $\chi^2_N$-analysis on optimized events at both the total cross-section level and the most sensitive differential distribution level.\footnote{To enhance sensitivity to the presence of ALPs in the \texttt{NC} (\texttt{CC}) processes, with $a\to\gamma\gamma$, we employ as our key observable the azimuthal angle difference between the scattered leptons $\Delta\Phi_{e^+e^-}$ (between the missing transverse momentum and one of the photons, $\Delta\Phi_{E_T^{\rm miss}\gamma}$). Since the $aVV$ vertex depends on two independent momenta ($p_{V_i, V_j}$), angular observables constructed solely from the decay photons cannot fully capture the underlying coupling structure as they depend only on one momentum $p_{a}$ combination. In contrast, the fermion lines (scattered $e^\pm$ or $E_T^{\rm miss}$) retain information about the mediator's momentum flow, hence used as the discriminating observables for the $\chi^2_N$ analysis. This ensures maximal sensitivity to the full Lorentz structure of the effective interaction.} For \texttt{CC} (\texttt{NC}), the most sensitive differential observable used in the $\chi^2_N$-analysis is the azimuthal angle difference $\Delta\Phi_{{E_T^{\rm miss}}\gamma_{2}}$ ($\Delta\Phi{e^+e^-}$), defined between the missing transverse energy and the sub-leading photon (scattered electron and positron). The $\chi^2_N$ is defined as:
\begin{align}
    \chi^2_N =&\, \sum_{k=1}^{n} \left(\frac{N_k(g_{ij}) - N_k^{\rm SM}}{\Delta N_k}\right)^2, \notag\\ 
    \Delta N_k =&\, \sqrt{N_k^{\rm SM} \left( 1 + \delta_s^2 N_k^{\rm SM} \right)},\label{eq:2}
\end{align}
where $N_k(g_{ij})$ represents the number of signal events in the $k^{\rm th}$ bin of a distribution, with a total of $n$ bins, $N_k^{\rm SM}$ is the corresponding number of background events. For our analysis, we assume $\delta_s = 5\%$ for a given luminosity $L$, with $L = 0.5$~ab$^{-1}$. The resulting 95\% confidence level (C.L.) constraints on $g_{ij}/f_a-m_a$ plane are shown in \autoref{fig:gij}.

In the Higgsstrahlung process, the ALP can be produced in association with a $\gamma$ or $Z$-boson via an $s$-channel $\gamma^*$ or $Z^*$ mediator. Unlike the $t$-channel fusion topologies, no $W$-mediated contribution is possible. The corresponding signal processes are $e^+ e^- \to a \gamma$ and $e^+ e^- \to a Z$, with the $Z$ boson decaying to $\ell^+\ell^-,\, jj$, or $\nu_\ell \bar{\nu}_\ell$ ($\ell = e,\,\mu$), and the ALP decaying as in the $t$-channel, $a \to \gamma\gamma$. The same simulation setup is used to generate these signals and the corresponding backgrounds, which include $3\gamma$, $2\gamma + \ell^+\ell^-$, $2\gamma + jj$, or $2\gamma + \nu_\ell \bar{\nu}_\ell$ as final states. The events are optimized using the required cuts on final-state photons, leptons, jets (ordered by $p_T$), and missing transverse energy. We further analyze sensitive observables such as $\Delta \Phi_{\gamma_1,\,\gamma_3}$, $\Delta \Phi_{\gamma_2,\,\gamma_3}$, $\Delta \Phi_{\gamma_1,\,j_2}$, $\Delta \Phi_{\gamma_2,\,j_1}$, $\Delta \Phi_{\gamma_2,\,\ell^+}$, and $\Delta \Phi_{E_T^{\rm miss},\,\gamma_2}$. A multi-bin analysis, similar to that employed for the $t$-channel process, is then performed, and the resulting 95\%~C.L. constraints are shown in \autoref{fig:chi_sChan}.

It is evident from \autoref{fig:gij} and \autoref{fig:chi_sChan} that the constraint from the $s$-channel process is better than that from the $t$-channel process in an $e^{+}e^{-}$ collider for the chosen setup.

\subsection{Muon anomalous magnetic moment:}
\label{const-amu}
%

The most recent high-precision determination of the muon anomalous magnetic moment has been consolidated in the latest update of the SM prediction~\cite{Aliberti:2025beg}, which yields 
$a_\mu^{\rm SM} = 116\,592\,033(62)\times 10^{-11}$ (530~ppb). 
When compared with the current experimental average, the difference is 
$\Delta a_\mu = a_\mu^{\rm exp}-a_\mu^{\rm SM} = 38(63)\times 10^{-11}$, 
indicating no statistically significant deviation between the SM and experiment at the present level of precision. However, the SM prediction faces tensions between lattice-based calculations of the hadronic vacuum polarization~\cite{Boccaletti:2024guq,Kuberski:2023qgx} and conservative data-driven estimates from $e^+e^- \to \text{hadrons}$ cross-section measurements~\cite{Keshavarzi:2019abf}. While ongoing improvements in the SM calculation may reduce the discrepancy with experiment, a resolution remains unconfirmed.

Various beyond the SM (BSM) scenarios have been proposed to explain the $\Delta a_\mu$ anomaly~\cite{Holst:2021lzm,Goyal:2022cmz,Sabatta:2019nfg,Athron:2021iuf,Capdevilla:2021rwo,Bharadwaj:2021tgp,Foldenauer:2018zrz,Drees:2021rsg}. In this work, we focus on contributions from ALPs interacting with charged leptons and electroweak gauge bosons. The leading ALP contributions to $\Delta a_\mu$ arise from both one-loop and two-loop diagrams, as shown in \autoref{fig:mug}. In particular, \autoref{fig:mug}~(a) and \autoref{fig:mug}~(c) provide negative contributions, while \autoref{fig:mug}~(b) can have either negative or positive contributions depending on the values of $g_{\mu\mu}$ and $g_{\gamma\gamma}$ over the $m_a$ range considered in this study. For light ALPs with $0.1~{\rm GeV} \leq m_a \leq 1~{\rm GeV}$, these effects have been studied in~\cite{Marciano:2016yhf,Keung:2021rps}. Extending the analysis to higher masses, \autoref{fig:g-2}~(a) shows the $1\sigma$ region satisfying $\Delta a_\mu$ in the $(g_{\gamma\gamma}/f_a,\, g_{\mu\mu}/f_a)$ plane for $m_a = 10$, 50, and 100~GeV. 

\subsection{Relic-density:}
\label{const-rel}

The Planck collaboration has precisely measured the DM relic density, reporting $\Omega_{\text{DM}} h^2 = 0.1198 \pm 0.0012$~\cite{Planck:2018vyg}. In ALP-portal models, DM candidates can achieve the observed relic density through freeze-out~\cite{Dolan:2017osp,Darme:2020sjf}, particularly for ALPs in the keV to GeV range and light DM masses $m_\chi < m_a$, where stringent constraints from collider, beam-dump, and astrophysical observations apply~\cite{Bauer:2017ris,Chang:2018rso,Mimasu:2014nea}. Recently, photon-lepton specific ALPs have also been proposed to address leptonic anomalous magnetic moments in the $0.1$–$1$~GeV mass range~\cite{Marciano:2016yhf,Keung:2021rps}.

In our framework, $\chi$ is a fermionic DM particle that interacts with the ALP and remains in thermal equilibrium in the early universe through annihilation channels such as
\begin{align}
\chi\bar{\chi} \to \gamma\gamma,\ \ell\bar{\ell},\ Z\gamma,\ aa,
\end{align}
with the $\gamma\gamma$ channel often dominating the cross section. The relic abundance is primarily determined by the thermally averaged annihilation cross section $\langle \sigma v \rangle$, which depends sensitively on the mass hierarchy between $\chi$ and the ALP.

For ALP masses in the range $m_a \in [1, 100]~\mathrm{GeV}$, two regimes emerge. In the non-resonant regime ($m_\chi < m_a/2$), the channel $\chi \bar{\chi} \to a a$ is kinematically forbidden, and DM annihilates into SM particles through off-shell ALP exchange, with the cross section scaling as $\langle \sigma v \rangle \sim m_\chi^2 / f_a^4$. In the resonant regime ($2 m_\chi \sim m_a$), the cross section experiences a Breit-Wigner enhancement via the $s$-channel resonance, significantly reducing the relic abundance. For $m_\chi > m_a$, the channel $\chi \bar{\chi} \to a a$ opens up and can dominate, especially when the ALP has sizable decay modes to photons. Constraints from CMB, and indirect detection via gamma rays from \emph{Fermi}-LAT and H.E.S.S., are particularly stringent for light dark matter below the GeV scale~\cite{Planck2018,Slatyer2013,FermiLAT2015,HESS2018,Proceedings:2012ulb,Planck:2013win}.

For the numerical evaluation of the relic density, we use the package \texttt{MicrOMEGAs-v6}~\cite{Alguero:2023zol}. Our parameter scan selects points within the $1\sigma$ range of the measured $\Delta a_\mu$. We vary the DM mass in the range $1~\mathrm{GeV} \le m_\chi \le 50~\mathrm{GeV}$ and scan $C_{a\chi}/f_a$ over the interval $1$–$10~\mathrm{TeV}^{-1}$. In \autoref{fig:g-2}~(b), we show the dependence of the relic density on the mass ratio $m_a/m_\chi$, using parameter choices that simultaneously satisfy the $\Delta a_\mu$ anomaly. For a given $m_\chi$, two distinct values of $m_a$ can yield the correct relic abundance while remaining consistent with $\Delta a_\mu$ constraints.

\begin{figure}[t]
    \centering
    {\includegraphics[width=1.0\linewidth]{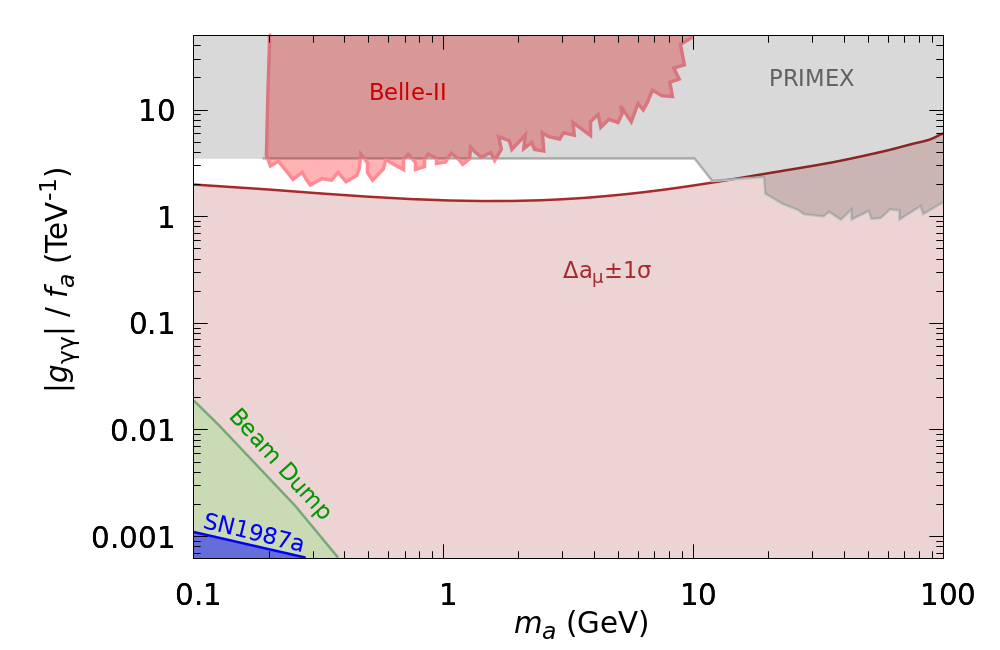}}
    \caption{Allowed parameter space in the $g_{\gamma\gamma}/f_a$–$m_a$ plane. The {\it brown} band represents the $1\sigma$ region favored by the $\Delta a_\mu$ anomaly, for which the entire parameter space also satisfies the dark matter relic density and the experimental measurement of $\mu_{Z\gamma}^{\rm exp}$ within $2\sigma$. A uniform bound of $|g_{Z\gamma}|/f_a < 0.7$~TeV$^{-1}$, derived from Higgs signal strength measurements~\cite{ATLAS:2020qcv,CMS:2022ahq,ATLAS:2023yqk}, is imposed across the full $m_a$ range. Additional constraints from PRIMEX~\cite{Aloni:2019ruo}, Belle-II~\cite{Belle-II:2020jti}, beam dump experiments, and SN1987A~\cite{Buen-Abad:2021fwq,Agrawal:2021dbo} are shown for comparison. The region with $m_a < 1$~GeV is included for illustration, where the $\Delta a_\mu$ anomaly is satisfied within $1\sigma$ and $\mu_{Z\gamma}^{\rm exp}$ within $2\sigma$, but relic density constraints are not applied. The $\Delta a_\mu$ constraint on $g_{\gamma\gamma}$ shown here is derived under the assumption of large $g_{\mu\mu}$ values, which excludes $|g_{\mu\mu}| = 0$ (see \autoref{fig:g-2}~(a)). For smaller $g_{\mu\mu}$, $\Delta a_\mu$ becomes negligible for all $g_{\gamma\gamma}$ values, while the remaining parameter space continues to satisfy the relic density and Higgs decay constraints.}
    \label{fig:hzy}
\end{figure}

\begin{figure*}[h!]
\centering
\subfloat[\label{fig:7a}]{\includegraphics[width=.50\textwidth]{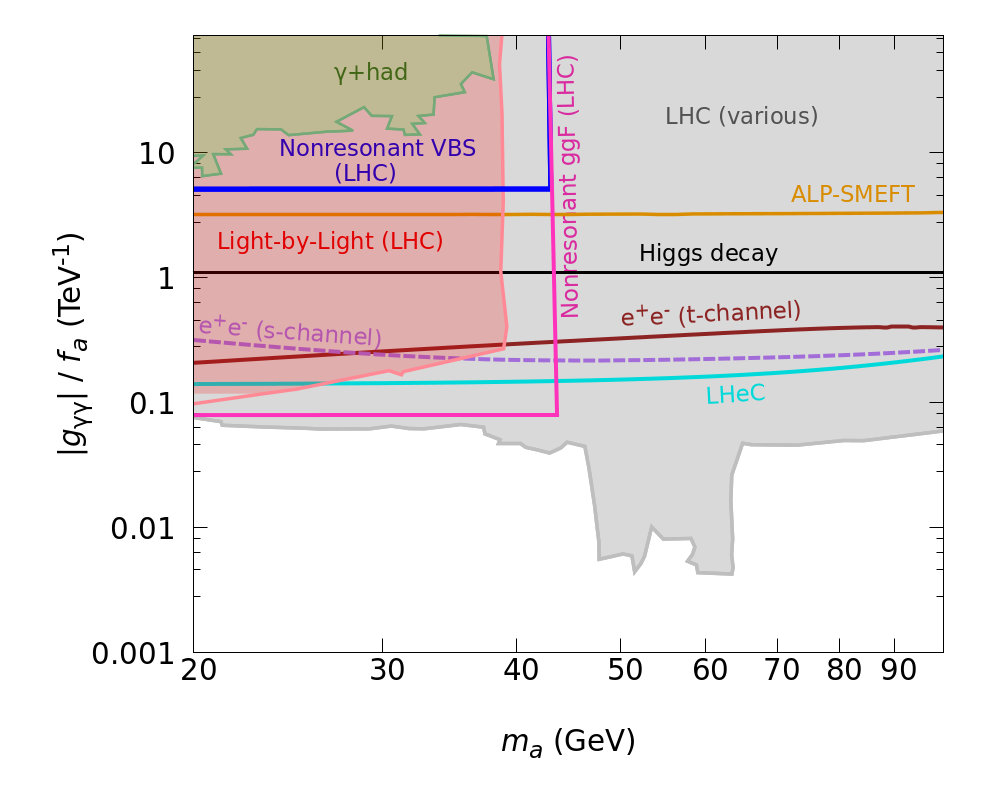}}
\subfloat[\label{fig:7b}]{\includegraphics[width=.50\textwidth]{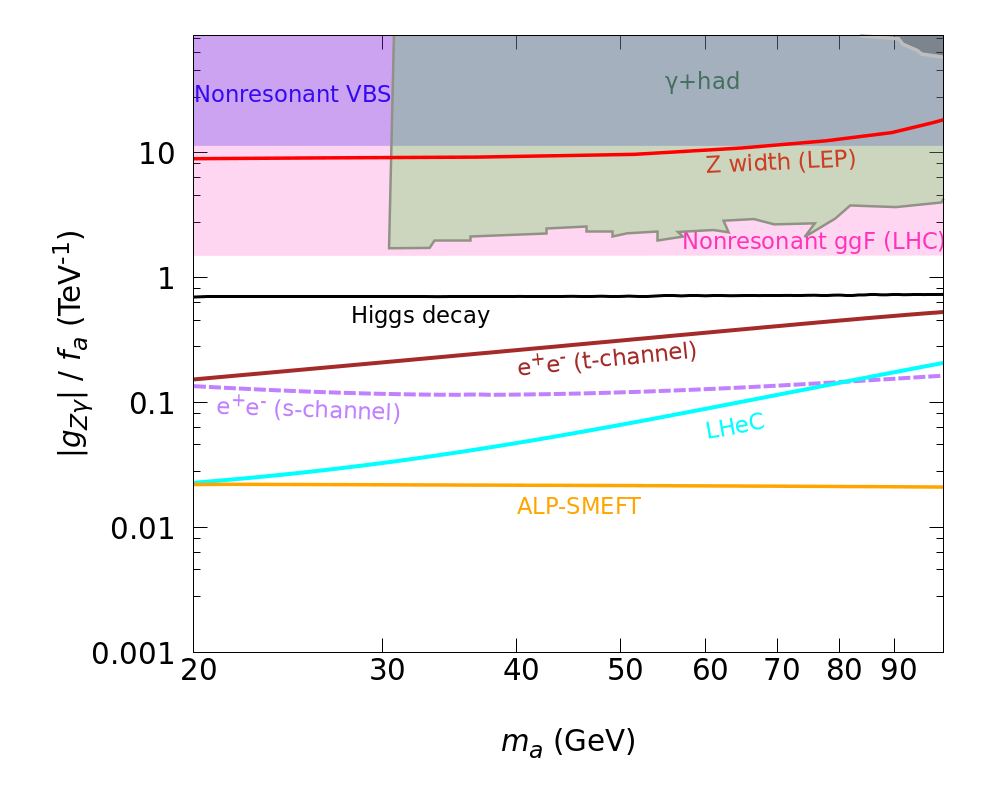}}\\
\subfloat[\label{fig:7c}]{\includegraphics[width=.50\textwidth]{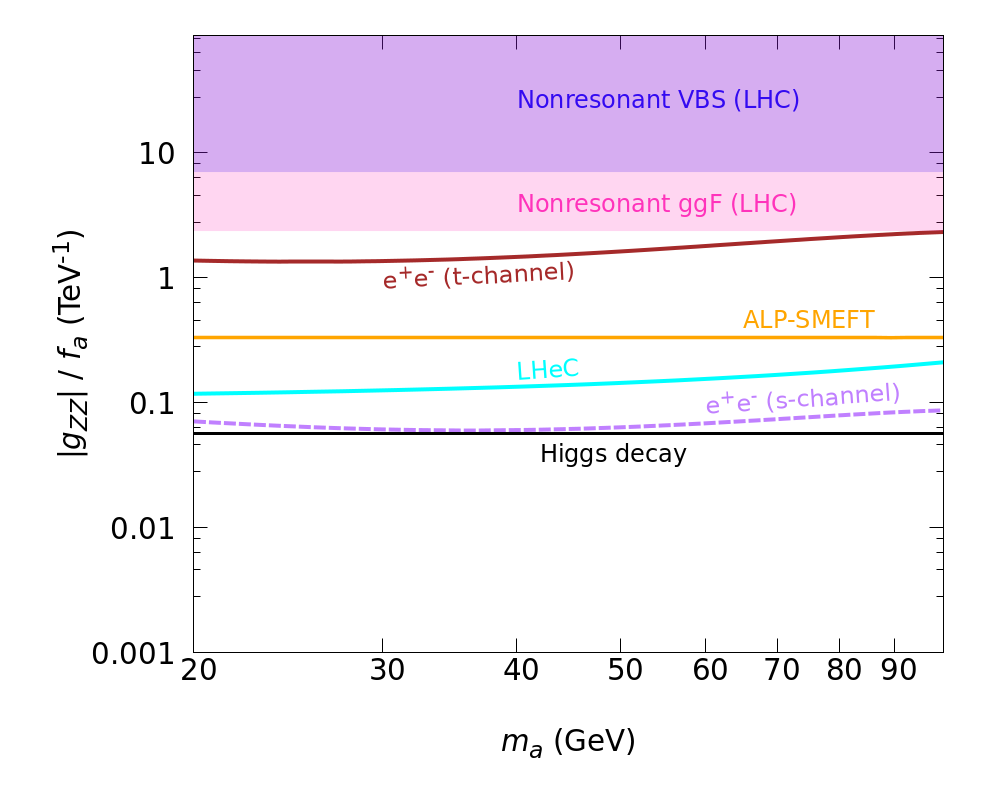}}
\subfloat[\label{fig:7d}]{\includegraphics[width=.50\textwidth]{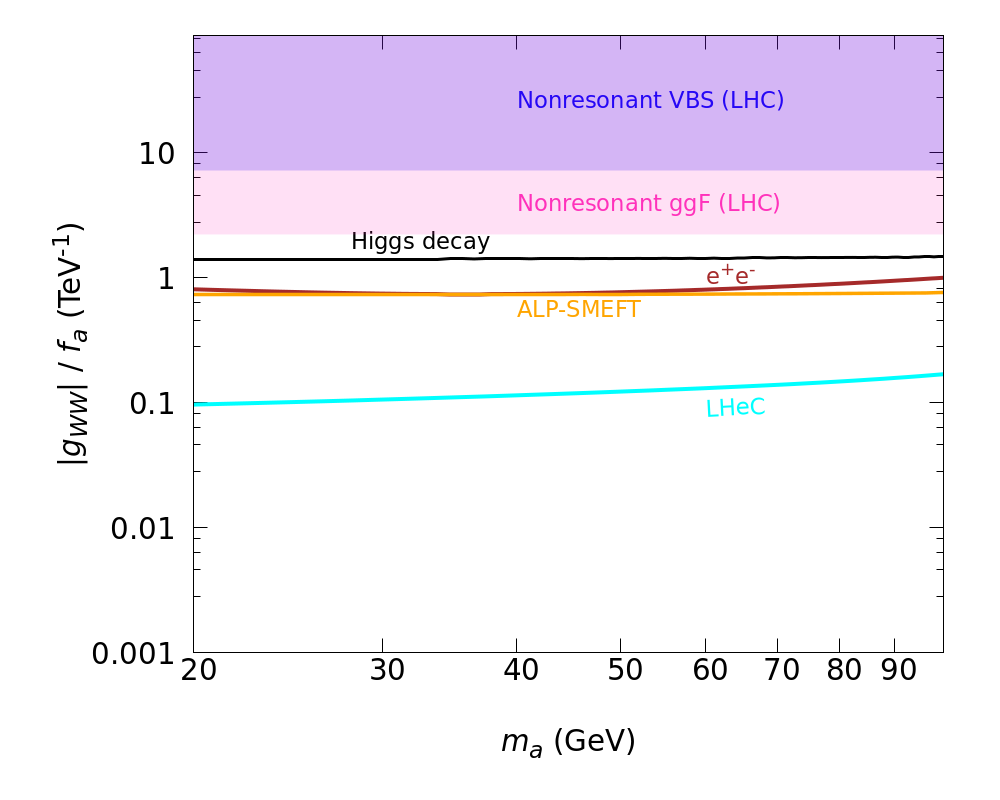}}
\caption{The 95\% C.L. exclusion contours in the $|g_{ij}|/f_a-m_a$ plane obtained from a multiple-bin $\chi^2$ analysis at $e^+e^-$ collider with $L = 0.5$~ab$^{-1}$. Limits from LHC, $\gamma$+hadron searches, $Z$-width at LEP, and ALP-SMEFT interpretations are shown for comparison~\cite{L3:1992kcg,Jaeckel:2012yz,Mariotti:2017vtv,CMS:2018erd,ATLAS:2020hii,Brivio:2017ije,Craig:2018kne,Gavela:2019cmq,Carra:2021ycg,CMS:2021xor,Bonilla:2022pxu,Alonso-Alvarez:2018irt,Biekotter:2023mpd,Bauer:2017ris,Bauer:2018uxu,Biswas:2023ksj}. The solid {\it brown} and dashed {\it purple} line shows our result from $e^+e^-$ analysis; the solid {\it black} line corresponds to constraints from $\mu_{\gamma\gamma,\,Z\gamma}^{\rm exp}-$ based analysis under a conservative assumption. For $g_{Z\gamma}$, interference effects with $g_{\gamma\gamma}$ and $g_{ZZ}$ are accounted for; a standalone limit is shown for $g_{\gamma\gamma} = 0.1 = g_{ZZ}$. The LHeC limits with $L = 1$~ab$^{-1}$ are adopted from Ref.~\cite{Mosala:2023sse}.
}
\label{fig:final}
\end{figure*}

\subsection{Higgs signal strength:}
\label{const-higgs}
Recent measurements by the ATLAS and CMS collaborations report a combined $h \rightarrow Z \gamma$ signal strength of $\mu_{Z\gamma}^{\rm exp} = 2.2 \pm 0.7$~\cite{ATLAS:2020qcv,CMS:2022ahq,ATLAS:2023yqk}, deviating from the SM prediction $\mu_{Z\gamma}^{\rm SM}=1$ at the $1.9\sigma$ level. While two-loop QCD and electroweak corrections~\cite{Spira:1991tj,Gehrmann:2015dua,Buccioni:2023qnt,Bonciani:2015eua,Chen:2024vyn,Sang:2024vqk} refine the SM estimate, they remain insufficient to explain this excess. Due to their loop-induced nature, the $h \to \gamma\gamma$ (with $\mu^{\rm exp}_{\gamma\gamma} = 1.10 \pm 0.06$~\cite{ParticleDataGroup:2024cfk}) and $h \to Z\gamma$ decay channels are highly sensitive to new physics. In particular, ALP contributions at one loop~\cite{Aiko:2023nox} can significantly modify these rates via effective couplings $g_{\gamma\gamma}$ and $g_{Z\gamma}$, as illustrated in \autoref{fig:mug}~(d).

The modified Higgs signal strength due to BSM effects is defined as 
\begin{equation} 
\mu_{XY}^{\rm BSM} = \frac{\Gamma_{\rm BSM}(h \rightarrow XY)}{\Gamma_{\rm SM}(h \rightarrow XY)}, 
\end{equation} 
where $XY = \gamma\gamma$ or $Z\gamma$, and $\Gamma_{\rm BSM}$ includes ALP-induced loop contributions. To constrain the ALP couplings, we construct a $\Delta\chi^2_\mu$ estimator, \begin{equation} 
\Delta\chi^2_\mu = \left( \frac{\mu_{XY}^{\rm BSM} - \mu_{XY}^{\rm exp}}{\Delta\mu_{XY}^{\rm exp}} \right)^2, 
\end{equation} 
and apply a $2\sigma$ bound, taking $\Delta\chi^2 = 4.0$ for single-parameter fits and $\Delta\chi^2 = 6.18$ for two-parameter fits.

For each ALP mass $m_a$ in the range 1–100~GeV, we first determine the values of $g_{\gamma\gamma}/f_a$ vs $m_a$ that satisfy the $\Delta a_\mu$ anomaly within $1\sigma$. Independently, we compute the allowed $g_{Z\gamma}/f_a$ range using the measured $\mu_{\gamma\gamma}^{\rm exp}$, applying a one-parameter $\Delta\chi^2_\mu$ test with a $2\sigma$ constraint. Combining these results, we scan over the $(g_{\gamma\gamma}/f_a,\, g_{Z\gamma}/f_a)$ plane and evaluate the consistency of each parameter point with the measured $\mu_{ Z\gamma}^{\rm exp}$ using the two-parameter $\Delta\chi^2_\mu$ test. For this calculation, the cutoff scale is set to $4\pi f_a$.

In \autoref{fig:hzy}, we show the allowed parameter space in the $g_{\gamma\gamma}/f_a-m_a$ plane. The {\it brown} band corresponds to the $1\sigma$ region favored by the $\Delta a_\mu$ anomaly, for which the entire parameter space simultaneously satisfies the dark matter relic density and the experimental measurement of $\mu_{Z\gamma}^{\rm exp}$ within $2\sigma$. Notably, for the $g_{Z\gamma}$ coupling, a uniform bound of $|g_{Z\gamma}|/f_a < 0.7~\text{TeV}^{-1}$ is observed across the entire mass range of $m_a$, and this constraint is therefore imposed as an additional bound. The coupling $C_{\mu\mu}/f_a$ is of order $\mathcal{O}(10^{1})~\mathrm{TeV}^{-1}$.

\section{Discussion and conclusions}
\label{diss}
The objective of this article was to present a coherent two-step framework to constrain Axion-Like Particle (ALP) couplings. First, we obtained independent bounds on the couplings $g_{\gamma\gamma}$, $g_{ZZ}$, $g_{Z\gamma}$, and $g_{WW}$ using projected sensitivities at an $e^+e^-$ collider with $\sqrt{s} = 250~\text{GeV}$ and integrated luminosity of $0.5~\text{ab}^{-1}$ for $m_a \in [20, 100]$~GeV. In the second step of our analysis, rather than interpreting the muon anomalous magnetic moment as an indication of new physics, we impose the latest determination of $\Delta a_\mu$ as a stringent consistency requirement on the ALP parameter space. We therefore identify the region in which the ALP contribution to $\Delta a_\mu$ remains compatible with the experimental value at the $1\sigma$ level, and further require this region to simultaneously satisfy the observed dark matter relic abundance and the Higgs signal strengths $\mu_{\gamma\gamma}$ and $\mu_{Z\gamma}$ within $2\sigma$. This combined set of observational constraints leads to substantially tighter bounds on the effective ALP–photon coupling ratio $g_{\gamma\gamma}/f_a$, particularly across the mass range $1~\text{GeV} \leq m_a \leq 100~\text{GeV}$. The resulting limits significantly reduce the allowed parameter space compared to those obtained from individual constraints alone, highlighting the complementarity between precision Higgs measurements, dark matter phenomenology, and the updated muon $g-2$ determination. The constraints shown in \autoref{fig:gij}, \autoref{fig:chi_sChan} and \autoref{fig:hzy} are obtained from two complementary analyses. Both analyses allow limits of order ${\cal O}(10^{-1})~\text{TeV}^{-1}$ on $g_{\gamma\gamma}/f_a$, assuming $f_a = 1~\text{TeV}$. 

To compare the obtained limits, the 95\% C.L. contours in the $|g_{ij}|/f_a-m_a$ plane are shown in \autoref{fig:final}. The solid {\it brown} (dashed \textit{purple}) line corresponds to constraints from the $e^+e^-$ collider analysis via the $t$-channel ($s$-channel) process, while the solid {\it black} line represents the most conservative bounds derived solely from the Higgs signal strength $\mu_{\gamma\gamma,\, Z\gamma}^{\rm exp}$ analysis. To extract limits on $g_{ZZ}$, $g_{Z\gamma}$, and $g_{WW}$ from $g_{\gamma\gamma}$ using the $\mu_{\gamma\gamma,\, Z\gamma}^{\rm exp}$ analysis, we employ the relations~\cite{Mosala:2023sse}:
\begin{equation}
g_{Z\gamma} = g_{WW} - s_w^2\, g_{\gamma\gamma}, \quad
g_{ZZ} = (c_w^2 - s_w^2)\, g_{WW} + s_w^4\, g_{\gamma\gamma}.
\end{equation}
The constraints on $g_{\gamma \gamma}/f_a$ obtained from the $t$- and $s$-channel processes in an $e^+e^-$ collider are comparable, while those on $g_{Z \gamma}/f_a$ and $g_{ZZ}/f_a$ derived from the $s$-channel are significantly stronger than the corresponding $t$-channel limits. At 95\% C.L., the $s$-channel process at an $e^+e^-$ collider yields stronger constraints on $g_{\gamma\gamma}/f_a$, reaching the level of ${\cal O}(10^{-1})~\text{TeV}^{-1}$, compared to those derived from Higgs decay measurements $h \to \gamma\gamma$ and $h \to Z\gamma$. Since the $g_{Z\gamma}$ coupling interferes with $g_{\gamma\gamma}$ and $g_{ZZ}$, we fix $g_{\gamma\gamma} = 0.1$ and $g_{ZZ} = 0.1$ for an isolated comparison. Under this setup, the $s$-channel $e^+e^-$ analysis provides bounds on $g_{Z\gamma}/f_a$ at ${\cal O}(10^{-1})~\text{TeV}^{-1}$, slightly stronger than those from Higgs data. Similarly, for $g_{ZZ}/f_a$, both Higgs and $s$-channel $e^+e^-$ results constrain it to the level of ${\cal O}(10^{-1})~\text{TeV}^{-1}$. The $g_{WW}/f_a$ limits from both sources lie around ${\cal O}(10^0)~\text{TeV}^{-1}$.

It is notable that employing a multi-bin analysis across several sensitive differential observables yields significantly stronger constraints than those reported in reference~\cite{Bauer:2018uxu}. Interestingly, as shown in \autoref{fig:hzy}, the PRIMEX experiment also constrains the parameter space in the range $15~\text{GeV} < m_a < 100~\text{GeV}$.

These findings demonstrate that precision measurements at future $e^+e^-$ colliders can probe ALP couplings with a sensitivity comparable to or exceeding that of current Higgs signal strength constraints. In particular, the collider setup offers complementary and, in some cases, stronger limits on $g_{ij}/f_a$, especially for $g_{\gamma\gamma}$ and $g_{Z\gamma}$. The coherent two-step approach outlined here provides a robust framework to test ALP scenarios motivated by the $\Delta a_\mu$ anomaly, dark matter relic abundance, and Higgs phenomenology. These constraints are obtained from full detector-level simulations with a conservative 5\% systematic uncertainty, reinforcing the robustness of the projections. This underscores the role of $e^+e^-$ machines as precision tools for ALP coupling reconstruction across a broad mass range.


\bibliography{example}
%
%

\end{document}